# Performance Evaluation of Infrared Image Enhancement Techniques

Rania Gaber[1], AbdElmgied Ali[2], and Kareem Ahmed[3]

[1]Computer Science department, NUB University, Egypt, raniagabeer@yahoo.com
[2]Computer Science department, Minia University, Egypt, abdelmgeid@yahoo.com
[3]Computer Science department, Beni-Suef University, Egypt, kareem_ahmed@hotmail.co.uk

**ABSTRACT**

Infrared (IR) images are widely used in many fields such as medical imaging, object tracking, astronomy and military purposes for securing borders. Infrared images can be captured day or night based on the type of capturing device. The capturing devices use electromagnetic radiation with longer wavelengths. There are several types of IR radiation based on the range of wavelength and corresponding frequency. Due to noising and other artifacts, IR images are not clearly visible. In this paper, we present a complete up-to-date survey on IR imaging enhancement techniques. The survey includes IR radiation types and devices and existing IR datasets. The survey covers spatial enhancement techniques, frequency-domain based enhancement techniques and Deep learning-based techniques.

## 1. INTRODUCTION

Image enhancement is a method of converting an image from its original state to a better one. Smoothing, noise reduction, edge detection, and other methods of enhancement are available. In essence, image enhancement is the first step in processing an image to improve its visual quality by reinforcing edges and smoothing flat areas of an input image[3]. Image enhancement is the process of enhancing the interpretability or perception of information in images by viewers while also producing better input for other automated image processing processes. The main goal of image enhancement is to change the characteristics of an image to make it more suited for a specific activity and observer. One or more image characteristics are changed throughout this operation. A task's choice of qualities and how they are updated are unique to that task. Furthermore, observer-specific elements such as the human visual system and the observer's experience will bring a significant amount of subjectivity into the selection of picture enhancing methods. There are a variety of ways that may be used to improve a digital image without ruining it.

The wavelength of Infrared (IR) light is longer than visible light; it ranges from the nominal edge of visible red light at 0.74 micrometers to 300 micrometers. These wavelengths correspond to a frequency range of around 1 to 400 THz and encompass the majority of thermal radiation emitted by objects at ambient temperature. When molecules vary their rotational-vibrational movements, IR light is released or absorbed microscopically. In many applications, images acquired through various infrared (IR) instruments are warped due to atmospheric aberration, which is caused by atmospheric fluctuations and aerosol turbulence[1]. According to International Commission on Illumination (CIE), the Infrared radiation is divided into three main bands: the first band is called IR-A where the infrared radiation has a wavelength from 700 nanometers (0.7 micrometers) to 1400 nanometers (1.4 micrometers), and a frequency between 215 terahertzes and 430 terahertzes. The second band is called IR-B where the infrared radiation has a wavelength from 1400 nanometers (1.4 micrometers) to 3000 nanometers (3 micrometers), and a frequency between 100 terahertzes and 215 terahertzes. The third band is called IR-C where the infrared radiation has a wavelength from 3000 nanometers (3 micrometers) to 1 millimeter (1000 micrometers), and a frequency between 300 gigahertzes and 100 terahertzes.

Modern approaches classify IR radiation into four categories: (1) near Infrared (NIR); (2) short wavelength Infrared (SWIR); (3) mid-wavelength Infrared (MWIR); (4) long wavelength Infrared (LWIR); and (5) Far infrared (FIR). The details and some examples of each type is as follows:

1. Near Infrared: this category contains light radiations with wavelengths between 0.75 micrometer and 1.4 micrometer. It is corresponding to frequency between 214 terahertz and 400 terahertz. Some examples of this category are night vision devices such as night vision goggles and Near-infrared spectroscopy.
2. Infrared with a short wavelength: this category contains light radiations with wavelengths between 1.4

micrometer and 3 micrometers. It is corresponding to frequency between 100 terahertz and 214 terahertz.

3. Infrared with a mid-wavelength: this category contains light radiations with wavelengths between 3 micrometer and 8 micrometers. It is corresponding to frequency between 37 terahertz and 100 terahertz. This band is often known as intermediate infrared (IIR) or thermal infrared.
4. Infrared with a long wavelength: this category contains light radiations with wavelengths between 8 micrometer and 15 micrometers. It is corresponding to frequency between 20 terahertz and 37 terahertz. This is the "thermal imaging" zone, where sensors can capture a completely passive picture of the outside world using only thermal emissions without any external light source.
5. Far infrared (FIR): this category contains light radiations with wavelengths between 15 micrometer and 1000 micrometers. It is corresponding to frequency between 0.3 terahertz and 20 terahertz. An example of this category is the Far-infrared laser.

Infrared images captured by various infrared (IR) instruments. Because of atmospheric changes and aerosol turbulence, images are warped in various applications due to atmospheric aberration. In recent years, as the cost of IR cameras has decreased, the number of applications based on IR images has expanded. In domains like as the military, medicine, and industry, many new applications of the infrared image have been created and deployed. Long-range surveillance is one of the most significant applications; nevertheless, IR images obtained at long range usually have low contrast, low brightness, and low magnification of hot objects of interest. Furthermore, low spatial frequency is an intrinsic restriction that impacts the quality of IR images and hampers their implementation. This is due to the IR optical system's diffraction effects, which cause the spatial frequency of IR images to be lower than that of visible images. In IR images, edge and texture information about objects is less defined than in visible images. As a result, in order for long-range surveillance to be effective, the photos must be processed to increase their quality. Image preprocessing is required to improve image quality and increase the acceptance of IR-based applications. The major goals of IR image preprocessing are to improve contrast enhancement and high spatial frequency enhancement[2].

Image Enhancement techniques can be divided into three main categories: (1) Spatial domain-based techniques; (2) Frequency domain-based techniques; and (3) Deep-learning based techniques. Image enhancement in the spatial domain [4]is accomplished by changing the image's pixel values, commonly known as point processing. Infrared image enhancement has grown increasingly difficult as a result of the examination of night vision cameras, as it provides an image with limited visibility and low contrast. The image produced by this approach is more like to how a human would see the scene. Various approaches are performed to the Fourier transformed image in the frequency domain to adjust its brightness, contrast, grey level distribution, and remove impulse noise. To achieve a smoother image, high frequency contents are removed, and low frequency contents are removed to achieve a sharper image. Deep-learning based approaches uses CNNs to train a network for performing IR image enhancement, then deploy the pretrained CNN to enhance new IR images.

**2. SPATIAL DOMAIN BASED ENHANCEMENT TECHNIQUES**

The spatial domain is used to define the actual spatial coordinates of pixels within an image, so when we use this term in the image enhancement business, we are talking about things like equalization, smoothing, and sharpening. [5]this paper includes information on spatial domain techniques for image enhancement, with particular reference to point processing methods and histogram processing. The main goal of image enhancement is to process an image such that the result is more suitable for a particular application than the original image. Digital image enhancement techniques provide a wide range of options for enhancing image quality. The imaging modality, mission at hand, and viewing conditions all affect the effective selection of such techniques.

The image pixels are directly addressed by spatial domain approaches. To accomplish the desired improvement, the pixel values are altered. Spatial domain techniques such as logarithmic transformations, power law transforms, and histogram equalization are based on direct pixel manipulation. Spatial approaches are very useful for changing the grey level values of individual pixels and, as a result, the overall contrast of the image. However, they frequently improve the entire image in a uniform manner, which yields poor outcomes in many circumstances. Edges and other needed information cannot be properly enhanced selectively. Techniques like histogram equalization are effective in many images. When operations are done on a single pixel, point processing operations (Intensity transformation function) are the simplest spatial domain operations. The processed image's pixel values are determined by the original image's pixel values[6]. The calculation

$$g(x,y) = T(f(x,y)) \quad (1)$$

Where T is the grey level transformation in point processing, can be used to calculate it. T is the only variable that changes in all spatial domain approaches. The previous equation can also be written in this way $S = T(r)$ where T is the transformation that converts pixel value r to pixel value s. Because we are only dealing with grey scale digital images, the results of this transformation are mapped back into the grey scale range. As a result, the findings are mapped back into the range [0, L-1], where L=2k is the number of bits in the image being processed, and k is the number of bits in the image being processed. We will only look at grayscale images here[7].

The first type of spatial domain-based enhancement techniques is the linear transformation, which is also called image negative. In this transformation, the pixels' values are inverted. Negative transformation is used to calculate image negatives when the intensity level is between [0, L-1]. The formula

$$s = l - 1 - r \quad (2)$$

is used to represent it. The intensity level of an image is reversed or inverted to create a photographic negative. If the darker sections of an image are more prominent and larger, we can use this strategy to improve the grey or white information paired with the darker areas of the image. In image negative, dehaze algorithm approaches and inverting input poor lighting video are applied. These algorithms are used to improve LCD displays and videos of poor quality[8]. Image Negatives, in which the grey level values of the pixels of an image are inverted to create a negative image, are one of four types of point processing procedures. Consider an 8-bit digital image of size $M \times N$, then subtract each pixel value from 255 as $g(x, y) = 255 - f(x, y)$ for $0 \leq x < M$ and $0 \leq y < N$. In normalized grayscales, the equation is expressed as $s = 1.0 - r$[6].

The second type of spatial domain-based enhancement techniques is the logarithmic transformation which is performed directly on gray scale values according to equation 3

$$s = c \, log \, (1 + r) \quad (3)$$

Where c stands for constant which is scaling factor to determine amount of enhancement. The image's darker pixel values are extended by compressing the values in the higher levels when employing this log transformation. The operation is carried out in reverse order for the inverse log transformation function. The major properties of log transformation include compression of dynamic range values in an image by introducing huge variations in the pixel value. Many researches applied the log reduction zonal magnitude technique to extend darker areas in IR images. Security surveillance applications are where these techniques are most commonly deployed[8-10].

The third type of spatial domain-based enhancement techniques is the power transformation, which is also defined as gamma correction. It is defined as follows:

$$s = c \, r^\gamma \quad (4)$$

This is the fundamental form of power law transformation; this strategy was employed where c and are $\gamma$ both positive constants. By applying power law curves with fractional values of ɣ the input value, which is narrow and has a darker range, the input value is transferred to the output value, which has a wider range. If higher level input values are provided, the process is performed in reverse order. Capturing an image, responding to its display, and reproducing the image are all examples of applications that use the power law technique. In general, the exponent of a power law equation is written as gamma. The process or procedure for rectifying or correcting the power law response is known as gamma correction[4, 8, 10].

The fourth type of spatial domain-based enhancement techniques is grayscale segmentation, also known as intensity level slicing or thresholding. When a specific range of grey levels becomes prominent and needs to be applied to enhance an image, i.e., to suppress or keep some important elements, this approach is used. Although it is similar to thresholding, it differs slightly in how it works. This method is the most commonly used to analyze most health and satellite photos for further investigation. It's especially useful for correcting and displacing objects like flaws in X-ray images and water masses in satellite imagery. The strategy to level slicing responsibility, regardless of variation, is based on two key ideas. This method involves displaying greater values for all grey levels in the region of interest and lower values for all other grey levels. The focus is on the object of interest rather than the background or other factors in this situation. There are an alternative way of this transformation, which brightens the desired gray-level range while preserving the other intensity levels[13].

A special type of intensity level slicing is Bit-Plane slicing which is widely used to enhance low contrast images. The image enhancement method frequently brightens all of the pixels in the input image. Bit plane slicing is commonly used to address this flaw. The image is sliced into eight bit-planes. Bits in bit plane 0 are referred to as least significant bits, whereas bits in plane 7 are referred to as most significant bits[14, 15]. Each pixel's intensity value can be represented by an 8-bit binary vector $v_i$ ($v_7, v_6, v_5, v_4, v_3, v_2, v_1, v_0$), where i is a value from 0 to 7 and each $v_i$ can have a value of 0 or 1.

The fifth type of spatial domain-based enhancement techniques is contrast stretching. One of the image enhancement techniques is contrast stretching, which includes editing an image to make it look better to human viewers. It's frequently used in post-production to change an image's contrast, dynamic range, or both. The goal of the contrast enhancement procedure is to alter the local contrast in different sections of a picture such that features in dark or bright areas can be seen by humans. Contract enhancement is commonly used to improve the visual representation of input images by altering original pixel values with a transform function of the form $g(x, y) = T[r(x, y)]$, where $g(x, y)$ and $r(x, y)$ are the output and input pixel values at image position, respectively. By stretching the range of grey level values to span a desired range of grey level values, this procedure improves contrast. Image intensity transformation or normalizing are other terms for this procedure. Let a, b represents f (x, y) minimum and maximum pixel values, while $c, d$ represent g $(x, y)$ minimum and maximum pixel values. By scaling each pixel in the original image value as $s = (r - c)(b - a)/(d - c) + a$, normalization can be achieved. Existing contrast enhancement approaches can be further classified into two categories: direct and indirect methods. Direct techniques attempt to improve a contrast measure by defining it. Indirect techniques, on the other hand, boost contrast without defining a specific contrast term by utilizing underutilized regions or the dynamic range[6]. The contrast stretching method is used to extend an image's dynamic range. The dynamic range of an image is the distance between the image's lowest and highest intensity values. Contrast Stretching is defined mathematically as,

$$Ì(X,Y) = \frac{d}{I_{MAX}-I_{MIN}} \times (I(X,Y) - I_{MIN}) + I_0 \quad (5)$$

Where $Ì(X, Y)$ is the new dynamic range image, d is the new dynamic range value, $I(X, Y)$ is the source image, $I_{MIN}$ is the input image's lowest intensity value, $I_{MAX}$ is the input image's highest intensity value, and $I_0$ is the new dynamic range's offset point for $Ì(X, Y)$. Although this transformation will produce a reasonable visual depiction of the original picture, some detail may be lost due to saturation and clipping, as well as poor visibility in the image's under-exposure parts[11, 12].

The sixth type of spatial domain-based enhancement techniques is histogram equalizations and histogram matching. Histograms are easy to calculate in software and may easily be implemented on low-cost hardware, making them a popular tool for real-time image processing. A digital image's histogram is a discrete function with intensity levels between [0, L-1]. The histogram function is defined as $h(k) = n_k$; Where, k represents the intensity value, $n_k$ is the number of pixels in the image with intensity k. Histograms are frequently normalized by the total number of pixels in the image. Assuming an M x N image, a normalized histogram or probability of occurrence of each gray scale [5, 16] is defined as $P(k) = n_k/MN$; Where $k = 0,1, \ldots, L-1$; and $P(k)$ is a probability estimate for the occurrence of grey level k. the sum of all probabilities for all gray scales is equal to 1. The grey levels are addressed by the histogram of an image. The use of histograms to determine whether a particular image is dark or light, low contrast or high contrast. It can be expressed as a discrete function, for example, in an image. It's utilized to improve the visual appeal of an image. Equalization of histograms is a typical approach for improving the appearance of images. Assume that we have a completely dark image. The visual detail is compressed towards the dark end of the histogram, and the histogram is skewed towards the lower end of the grey scale. The image would be much clearer if we could 'stretch out' the grey levels at the dark end to obtain a more consistently distributed histogram. Histogram Equalization (HE) is a technique for adjusting contrast in images by using the histogram. The goal behind this technique is to remap the scene's histogram to a histogram with a near-uniform probability density function. Equalization of the histogram redistributes the intensity distribution. If an image's histogram has a lot of peaks and valleys, it will still have peaks and valleys after equalization, but they will be moved. Histogram Equalization is a technique that increases contrast and aims to achieve a homogeneous histogram. Global Histogram Equalization (GHE), Adaptive Histogram Equalization (AHE), and Block-based Histogram Equalization are the three methods of histogram equalization (BHE). Each pixel is given a new intensity value based on the prior cumulative distribution function in Global Histogram Equalization (GHE). The original histogram of the grayscale image must be equalized before doing Global Histogram Equalization (GHE). The input image's cumulative histogram must be equalized to 255 by applying a new intensity value to it[19].

All of spatial domain transformations can be performed globally or local. Histogram equalizations and histogram matching methods used in the earlier were global. As a result, local augmentation is employed. Move the center of the square or rectangular neighborhood (mask) from pixel to pixel. Calculate a histogram of the points in each neighborhood for each neighborhood. Obtain the equalization/specification function for histograms. Map the grey level of a pixel in a neighborhood. It can calculate the next histogram using updated pixel values and the previous histogram. In the case of medical and satellite images, it is also vital to improve local information in order to achieve better results. Each and every pixel of an image has local information. By processing the neighboring pixels, histogram equalization and specification are more suitable for local information enhancements.

**3. FREQUENCY DOMAIN METHODS:**

Rather than manipulating the image itself, frequency domain techniques manipulate the image's orthogonal transform. Frequency domain techniques are well suited to image processing based on frequency content. The frequency domain image enhancement approaches are based on computing a 2-D discrete unitary transform of the image, such as the 2-D DFT, modifying the transform coefficients using an operator M, and then performing the inverse transform[6]. The plane of a two-dimensional discrete Fourier representation of an image is referred to as the frequency domain. When non-periodic functions may be written as the integral of sines and/or cosines multiplied by a weighting factor, the Fourier transform occurs. The Fourier transform defines the frequency domain, which has various uses in image processing such as image analysis, filtering, reconstruction, and compression. The DFT discrete Fourier transform of an image can be used to explain the principle of image filtering in the frequency domain. After applying the filter function, the inverse discrete Fourier transforms are performed. The integral transform of one function into another is known as the Fourier Transform. Fourier transform was invented by Jean Baptiste Joseph Fourier (1768-1830), a French mathematician and physicist. For many years, the FFT has played a significant role in image processing[21]. Assume that f(x) is a continuous function of the real variable x. F (u) is the Fourier Transform of $f(x)$:

$$F(u) = \int f(x)e^{-2\pi i u x} \, dx \quad (6)$$

Where $i = \sqrt{-1}$. The inverse Fourier Transform is defined as follows:

$$f(x) = \int F(u)e^{2\pi i u x} \, du \quad (7)$$

In case of discrete 2D discrete Fourier Transform, it is defined as:

$$F(k,l) = \frac{1}{MN} \sum_{m=0}^{M-1} \sum_{n=0}^{N-1} f[m,n]e^{-2\pi i \left(\frac{k}{M}m + \frac{l}{N}n\right)} \quad (8)$$

Where $l$ and $k$ are integers and $0 \leq l \leq N-1$ and $0 \leq k \leq M-1$. The 2D inverse Fourier Transforms is defined as:

$$f(m,n) = \sum_{k=0}^{M-1} \sum_{l=0}^{N-1} F[k,l]e^{2\pi i \left(\frac{k}{M}m + \frac{l}{N}n\right)} \quad (9)$$

In the frequency domain, the concept of filtering is easy to grasp. As a result, DFT-based frequency domain augmentation of image f (x, y) is possible. If the spatial extent of the point spread sequence h (x, y) is big, this is especially important in convolution theory.

To perform image enhancement in the frequency domain, the image is first converted to Fourier transform; then the resulting Fourier coefficients are filtered; finally, inverse Fourier transform is applied to restore the enhanced image. Filtering is the process, in which the image is modified to emphasize certain edges or objects; removing noise and artifacts; and blurring/deblurring certain regions. The filtering in the frequency domain is always performed for two basic operations; the first operation is smoothing; and the second operation is sharpening. The two operations are performed by removing low or high frequency components and maintain the other component. There are three main filters that are always applied to frequency domain: low-pass filters, high-pass filters; and band-pass filters. Low-Pass filter is also known as (smoothing or blurring) filter. It is used to smooth out images and reduce noise. The average of a pixel and all of its eight neighbors is calculated in the low-pass effect. Convolution that attenuates high frequency of an image while permitting low frequency through (leaving) is known as low-pass filtering. The High-Pass filter smooths the image and reduces noise, whereas the Low-Pass filter smooths the image and reduces noise[22]. There are three types of low-pass filters: Ideal low-pass filters, Butterworth low-pass filters, and Gaussian low-pass filters. The High-Pass filter, often known as the (Sharpening) filter, is used to sharpen images and highlight image details. High-Pass convolution, like Low-Pass convolution, calculates the average of a pixel and all of its eight neighbors, but it attenuates the image's low frequency by allowing high frequency to pass while modifying the Low frequency with convolution average[23]. A high-pass filter is one that allows high frequencies to pass through while attenuating sounds below the cut-off frequency. Sharpening is essentially a frequency domain high pass operation. There are three types of high-pass filters: Ideal High- pass filters, Butterworth High- pass filters, and Gaussian High- pass filters.

## 4. HOMOMORPHIC FILTERING

Homomorphic Filtering is a technique for improving images. It increased contrast and normalized brightness across an image at the same time. To reduce multiplicative noise, Homomorphic Filtering is utilized. Illumination and reflectance are indistinguishable in the frequency domain, but their approximate locations can be determined. Because illumination and reflectance are multiplicatively blended, the component is made additive by taking the logarithm of the image intensity, resulting in a frequency domain that is linear. Filtering in the log domain helps reduce illumination variations, which can be thought of as multiplicative noise[11]. In the log domain, homomorphic processing on images transforms the image into illumination and reflectance components. Low frequency components are those that have a gradual spatial change, and high frequency components are those that have a rapid spatial variation. It has the ability to reduce the brightness range and improve image contrast. To reduce non-uniform lighting without losing image features, log was applied individually to the brightness and reflectance components. Homomorphic filtering is widely used in image enhancement to improve the appearance of grayscale images by compressing intensity range which is represented by illumination; while performing contrast enhancement which is represented by reflection. The overall objective of Homomorphic filtering is to remove multiplicative noise; which is harder to be removed than additive noise. First the original image is decomposed into illumination and reflectance, as follows:

$$m(x,y) = i(x,y) \circ r(x,y) \quad (10)$$

Where m is an image, i is an illumination, and r is the reflectance. Then, we use the log function to transform the equation into two added terms instead of multiplied terms; to facilitate the calculations of Fourier transform; as follows:

$$\ln(m(x,y)) = \ln(i(x,y)) + \ln(r(x,y)) \quad (11)$$

Next, we apply the Fourier transform, as follows:

$$F(\ln(m(x,y))) = F(\ln(i(x,y))) + F(\ln(r(x,y))) \quad (12)$$

Then we perform the required low/high pass filtering, then performing the inverse Fourier transform followed by exponential transform to restore the enhanced image.

## 5. RETINEX THEORY

Land proposed the Retinex theory[24], which modifies the perceived object color and brightness model using the human visual system. Given that reflectance often describes an object's nature of reflection properties, we can use Retinex to obtain the true picture of an object. Many imaging applications, such as image enhancement, face recognition, and image retrieval, have used Retinex. How to estimate the illumination is a crucial step in Retinex theory. Jobson et al. proposed single-scale Retinex (SSR) and multi-scale Retinex (MSR) as modifications to Retinex theory. To stress a certain wavelength band of the image, SSR uses a Gaussian low-pass filter (LPF) and log operation. Using numerous linear LPFs with varied support areas, MSR produces an output image that is the weighted sum of the SSR output images[25].

Jobson et al. developed the first technique, SSR, in 1997, based on the center/surround Retinex. In SSR, the illumination layer is assumed to be the result of a Gauss transform over the image. It subtracts the logarithmic transform of both the image and illumination layer to get the output. In SSR, we'll be given a low-quality image to work with, and we'll have to improve it. As a result, we'll convolve the image with an appropriate filter created with the help of the surround function and treat it as an illumination. Then, using Retinex's method, we subtract the log of found illumination from the log of an input image to find the reflectance. Then, as an output, the outcome will be handled as an upgraded image. It's called single scale Retinex because it only employs one surround function to determine lighting by convolving images[26]. It can be expressed mathematically as:

$$R(m,n) = \log I_i(m,n) - \log[F(m,n) * I_i(m,n)] \quad (13)$$

Where $I_i$ denotes the image's distribution over the $i^{th}$ colour spectral band, and F(m, n) stands for the surround function, and R(m, n) stands for the output of the respective Retinex. After the surround function's convolution operation over the image, the logarithmic transformation is applied. The following is a mathematical depiction of the surround function:

$$F(m,n) = K * exp\left(\frac{-r^2}{c^2}\right) \quad (14)$$

The scalar value C can be referred to as either the surround space or Gaussian space constant. K was chosen based on the following factors:

$$r = \sqrt{(m^2 + n^2)} \quad (15)$$
$$\iint F(m,n) \, dm \, dn = 1 \quad (16)$$

The adjustment between dynamic range compression and total rendition is controlled by the Surround space constant. Using scales with smaller magnitudes can result in more dynamic range compression, while using kernels with big magnitudes can result in more color consistency.

At any given time, SSR can deliver entire rendition or dynamic range compression. SSR is unable to deliver both dynamic range compression and complete rendition simultaneously. Because it only employs a single scale to determine illumination, it may not be able to provide appropriate results for some photos if it only keeps one property of shrinking dynamic range or total rendition. As a result, a new strategy known as MSR was proposed to solve this (Multi scale Retinex). If the dynamic range of a scene is substantially greater than the dynamic range of the image recording device, unrecoverable information loss can occur. The MSR was created to address the shortcomings of the SSR. MSR mixes the quality of many surround spaces to provide a picture with good dynamic range and total rendition compression. To determine the illumination in MSR, it employs several surround functions on a single stage. We must obtain their log after computing the illumination from all different scales using the convolution procedure. Then remove the computed illumination from the log of an image for one scale, and repeat for every computed illumination. Then take the weighted average of all the results for all the scales to get the final improved image as an output. Multi scale Retinex is the addition of weighted outputs from several SSR scales. MSR can be expressed mathematically as:

$$R_{MSR} = \sum_{j=1}^{N} w_j \{ \log I_i(m,n) - \log[I_i(m,n) * M_j(m,n)] \} \quad (17)$$

Where I denotes the color band (i.e. R, G, B), N denotes the number of scales utilized, $M_j(m,n)$ denotes the surround functions, and $w_j$ denotes the scale's weighting factors. The surround function can be calculated using the following formula:

$$M_j(m,n) = K_j exp\left[\frac{-(m^2+n^2)}{\sigma^2_j}\right] \quad (18)$$

Where $\sigma^2$ is the Gaussian distribution's standard deviations.

After improving an image, MSR keeps the majority of the details. MSR is superior to SSR, but it is unable to provide the natural image that is central to the Retinex concept. MSR is capable of delivering results with both total rendition and dynamic range compression, however it does have a color sensitivity constraint. MSR is a great way to figure out how to improve a grayscale image, but it has challenges with color images. We can't tell whether the color obtained after processing the image is correct or not in MSR. MSR is experiencing difficulty with color sensitivity[26].

## 6. WAVELET MULTI-SCALE TRANSFORM

In recent years, wavelet analysis has shown to be a strong image processing method. The wavelet transform (WT) is the mathematical tool of choice when images must be seen or processed at numerous resolutions. The WT provides great insight into an image's spatial and frequency features, in addition to providing an economical and intuitive framework for the encoding and storage of multiresolution images. The image detail parts of a wavelet-transformed image are saved in the high-frequency parts, while the imagery constant part is recorded in the low-frequency part. The low-frequency element controls the dynamic range of the image because the imaging constant part determines the dynamic range of the image. To reduce the dynamic range, we attenuate the low-frequency component. When the low-frequency component is attenuated, however, the details must be lost. The image reconstructed using the inverse wavelet transform (IWT) contains more detail because some details are stored particularly well in the high-frequency sections[12]. The Spatial Enhancement methods work with the intensity values of the input image directly. Frequency domain approaches use DCT, DFT, and other image transformation techniques, as well as image frequency filtering. A hybrid approach is a combination of frequency and spatial methods (or) a strategy that combines frequency and spatial methods for better image enhancement. [27]A hybrid filter was created to remove more mixed (Gaussian and impulsive) noise from the image. The median filter was used to eliminate impulse noise, and the average filter was used to remove Gaussian noise. The approach was proposed as being reliable and simplifying.

## 7. CNN-BASED ENHANCEMENT TECHNIQUES

Bhattacharya et al. [49] presented a Convolutional Neural Network (CNN) for enhancement of maritime infrared images. they propose several CNNs to improve the resolution and reduce noise. They used one residual CNN for denoising to reduce graininess; and a super-resolution CNN to map low-resolution image to multi-scale high-resolution images.

Several attempts were made to deploy deep learning for enhancement of infrared image enhancement, but traditional CNNs such as residual architectures and encoder–decoder architectures, always produces bad results for infrared image enhancement. The training is very time consuming but not inclusive which results in bad performance in network architecture and enhancement accuracy. In [48], a fully convolutional neural network (CNN) is proposed for enhancing the contrast and details of a single infrared image, which is trained using only visible images. The basic idea of this proposed CNN is to incorporate the conditional generative adversarial networks into optimization framework to prevent background noise to be amplified. The proposed CNN was proved to enhance the IR image and produces a very good enhanced image with higher contrast and sharper details.

## 8. INFRARED DATASETS

There are many thermal IR image datasets which can be used as a benchmark for performance evaluation of different IR enhancement techniques. In [42], the authors presented a thermal imaging dataset for person detection. The dataset contains 7412 thermal images that are collected in different conditions and positions such as walking, running, and sneaking. They used long wave infrared in several conditions for the weather such as clear weather, foggy weather, and rainy weather. They also captured images at different distances between the camera and object. The dataset collects different ways for performing actions at different speeds.

In [43], The OSU Thermal Pedestrian Database is presented which is used for Person detection from thermal images. They used Raytheon 300D thermal sensor for collecting thermal images of Pedestrian intersection on the Ohio State University campus. The dataset contains 284 8-bit grayscale images of size 360 x 240 pixels. The ground truth is also provided for people that are clearly visible inside the image as list of surrounding boxes.

In [44], the CBSR NIR face dataset is presented for NIR eye detection and NIR face recognition. They used an NIR camera with active NIR lighting for capturing the images. There are 3940 NIR face images of 197 people; each image is 480x640 pixels, 8 bit, without compression. The images are divided into two sections, the gallery set and a probe set. In the gallery set, each person has eight images. In the probe set, each person has 12 images. There are also additional information containing image number, person number, and eye coordinates.

In [45], the Audio-Visual Vehicle (AVV) dataset is presented. The overall objective of the dataset is to detect moving vehicle under different challenging conditions such as occlusions, motion blur, and from different perspective views. They used Standoff long distance Laser Doppler Vibrometer (acoustic), Polytech LDV OFV505, HeNe laser 632 nm with two PTZ cameras, Canon VC-C50i to collect the samples. The dataset contains 961 sets of multimodal vehicles samples collected from two road: local road and a highway. Each sample consists of three files: an audio clip (mono 22.5kHz, 16 bit), an original image shot, and a reconstructed visual image. There are several objects that have been collected such as bikes, buses, motocycles, 2-door and 4-door sedans, different types of trucks and vans.

In [46, 47], the CSIR-CSIO Moving Object Thermal Infrared Imagery Dataset (MOTIID) is presented, which contains thermal infrared images for moving objects like pedestrians and vehicles. A thermal infrared camera is used to collect images from a height of 4 feet. The dataset contains 18 thermal sequences; containing images of size 640x480 pixels. The duration of these thermal sequences range from 4 to 22 seconds. Each sequence contains one or more moving objects entering and leaving.

## 9. PERFORMANCE EVALUATION

Because they are unaffected by lighting conditions, infrared imaging sensors are commonly used in urban traffic systems. However, due to hardware and imaging environment limitations, obtaining high-quality infrared (IR) images is difficult. With the traditional method, IR images always lack specific information, resulting in unsatisfactory IR image enhancing results. The visible (VIS) images, in contrast to the infrared (IR) photos, contain extensive information that may aid in improving the quality of the matching IR images. They suggest an effective strategy for enhancing IR photos using multi-sensor images in this study. According to Retinex theory, we first use an edge-preserving filter to deconstruct the IR and VIS images into illumination and reflectance components. Second, based on the correlation between the IR and VIS images, each region in the IR and VIS image is classified as related or unrelated. Finally, to improve the illumination component, an adaptive fuzzy plateau HE (AFPHE) is used, and an approach is used to improve the detail of the IR reflectance component using VIS images. The proposed method successfully improves the contrast and enhances the detail of IR images, according to experimental data.

Peak Signal to Noise Ratio (PSNR) represents the image's integrity, Structural similarity (SSIM) reflects how similar two images are, and the higher the value of these indexes, the greater the increased image quality.

and image sharpness and texture detail can be improved using the proposed method. When enhancing the contrast and detail information of the infrared image and improving the visual effect, this article can successfully improve the edge and contour detail of the image, and the target details are remarkable when the VIS clarity is better than the IR clarity. César[30], in this method yielded the following results on average for each iteration for the 450 photos. It improves contrast and signal-to-noise ratio while also increasing the level of detail. Furthermore, the brightness is better preserved. This technique has a longer computation time. As a result, it is the one that offers more details and maintains a higher brightness. The proposed strategy improves entropy while also preserving brightness and increasing contrast. When compared to state-of-the-art algorithms, the proposed technique was shown to be competitive. It's worth noting that the proposed method is the only one that enhanced the contrast of the original image for all input images. The final image, after using the proposed method, has a greater visual quality

| Technique | Enhanced image (contrast improvement) | AVERAGE PSNR/SNR | SSIM | AVERAGE running time (s) | Application |
|---|---|---|---|---|---|
| Wang[28] | - | 30.8847 | 0.4021 | - | Infrared fault detection. |
| Chen[29] | - | - | 0.7042 | - | Surveillance system. |
| César[30] | - | 25.455 | - | 12,609(MS) | infrared thermal image analysis, object recognition, people tracking |
| Kuang [31] | | 26.1337 | 0.9496 | | |
| Wang[32] | 0.07287 | - | - | 1.17 | network performance and range of application |
| Wan[33] | 7.2467 | - | - | 100.3737 | salient object detection, maximum power point tracking, travelling salesman problem, inverse radiative transfer |
| Fan[34] | 0.05803 | - | - | - | Target detection and recognition. |
| Wu[35] | 8.03 | - | - | - | numerous real-world applications |
| Qi[36] | 0.11 | - | - | 0.193 | Object detection and other infrared-based optical applications. |
| Yuan[37] | 0.432 | - | - | - | |
| Zhao[38] | 0.0606 | - | - | 5.0213 | Target detection and recognition. |
| Liang[39] | 57.6603 | - | - | - | The algorithm can be used in various applications to effectively enhance infrared images, especially details in the images |
| Bai[40] | 0.2224 | - | - | 54.032 | applications of infrared target detection and tracking |
| Lai[41] | 0.226 | - | - | - | is applied to nearly all infrared imaging based engineering applications |

**Table 1**: Performance evaluation of different IR image enhancement techniques

Table 1, demonstrates different results of infrared image optimization. Wang[28],The algorithm has a higher image enhancement performance index in PSNR, and the resulting enhanced image not only successfully warns people of potential danger, but also precisely locates malfunctioning. Chen [29], The structural similarity index, which is proposed, outperforms existing techniques. Image noise can be successfully removed while detail information is preserved,

than the original image. Kuang [31], Create a new refined convolutional neural architecture that generates visually appealing outcomes with more contrast and finer features than other network topologies. Because there are fewer infrared images, visible images are used for training. To ensure that the network trained on visible pictures can be applied to infrared images, proper training samples are generated. Experiments show that in terms of contrast and detail augmentation, our method beats existing picture enhancement algorithms. Wang[32], The subject of infrared image contrast enhancement has been thoroughly investigated in this study. MPSO algorithm with superior performance has been presented to enhance the contrast of infrared images. Wan[33], For IR image improvement, a PSO optimization-based local entropy weighted histogram equalization is proposed in this study. The proposed histogram is separated into fore- and background sections using a threshold that maximize the inter-class variance, allowing for enhanced foreground and background contrast in the equalization process. The PSO algorithm is then used to optimize the upper and lower thresholds of the histogram in order to avoid over-enhancement as much as feasible. Finally, the two sub-histograms are separately equalized, and the improved image is created. Our strategy improves fore-and background contrast, suppresses image noise, and avoids over-enhancement, according to both qualitative and quantitative testing. The PSO-based optimization entails several equalization iterations, resulting in a lengthy computation time. In reality, computing burden is a prevalent problem in all PSO-based algorithms that needs to be addressed immediately Fan[34], The targets are amplified and background clutter is removed from the dim infrared image. Subjective and quantitative evaluations show that the suggested algorithm performs well in improving contrast. Other algorithms obtain a smaller value than the proposed algorithm. This suggests that the suggested technique is the most effective at enhancing contrast in inferred images that are faint. Wu[35],The proposed approach efficiently increases the IR image's global and local contrast, as well as suppressing the image's noise. Several experiments are carried out to assess the proposed scheme's performance. The proposed system outperforms the previous methods in terms of both visual quality and objective evaluation, according to the results of the experiments. Qi[36], In this research, Qi use Cellular Automata to offer a new enhancement approach based on two priors. First, they use Cellular Automata to directly learn the gradient distribution prior from the photos. Second, they present a new gradient distribution error to encode structural information using Cellular Automata, taking into account the importance of image features. The system is straightforward to use, understand, and expandable to handle other vision tasks, as well as producing more accurate results. Yuan[37], The adaptive trilateral contrast enhancement (ATCE) method is provided as a contrast enhancement method. Unlike traditional approaches that emphasize the contrast difference between the foreground and background of an image to increase picture visual quality, the ATCE method takes a multidimensional approach that includes both image contrast and subtle image details augmentation. On the metric of enhancement by entropy, the quantitative experimental results reveal that ATCE greatly outperforms other available approaches. ATCE tries to improve on the traditional HE model by merging it with the concept of high-boost filtering to give both image contrast and detail improvement in a single multilayered process, resulting in a simple, flexible, and elegant solution Zhao[38], Based on multi-scale decomposition and saliency feature extraction, Zhao present a rapid IR image enhancing technique. In comparison to current methods for IR image enhancement, the suggested approach has significant advantages. To begin, a smooth based multi-scale decomposition is used to generate sub-images with distinct frequency components, allowing image information to be separated at multiple scales. Second, the local frequency-tuned technique was effective in extracting regions of interest, particularly target areas. It is critical and beneficial for target detection and recognition applications. Finally, they can improve target contrast by providing medium frequency components a lot of synthetic weight to accentuate the texture or profile of dim target regions. The results of the experiments show that the proposed strategy is both efficient and reliable. Liang[39], proposes and evaluates a new adaptive contrast enhancement algorithm based on double plateaus histogram equalization. The proposed algorithm is effective for enhancing infrared images in various scenes in real time by computing and updating upper and lower thresholds based on local maximums of non-zero histograms and the minimum grey interval. The technique with two adaptive thresholds has been shown to be capable of restricting background noise while boosting image details in image enhancement studies. The proposed algorithm can improve contrast enhancement by 2–10 dB when compared to traditional double plateaus histogram equalization. The proposed approach limited the background of an infrared image while also enhancing the features. The suggested algorithm may successfully enhance the contrast of infrared images, especially the details of infrared images, according to experimental results. Bai[40], An approach based on multiscale new top-hat transform is suggested to improve infrared images. The first step is to extract multiscale light and dark image regions using multiscale new white and black top-hat transforms. The recovered multiscale light and dark image regions are then used to create the final light and dark image regions. Finally, employing a power technique, the original infrared image is augmented by widening the contrast between the light and dark image regions. The proposed approach could significantly improve infrared images, particularly those with dark target regions. The proposed algorithm has a longer computation time than previous algorithms, but it has a good implementation and might be used in a variety of situations. Lai[41], For an infrared (IR) image, a scene-adaptive contrast enhancement technique based on quantitative measures is proposed. This approach enhances the raw IR image by first regulating the probability density function (PDF) of the raw image, and then applying an enhanced plateau histogram equalization method whose plateau threshold is controlled by the concavity of the regulated PDF. This method's parameter tuning strategy is based on finding the best parameters that correspond to the highest EME. The suggested approach's effectiveness is evaluated using a real infrared image, and the findings show that our method can successfully increase the dynamic range of the targets while suppressing the background and noise.